\documentclass[journal=jacsat,manuscript=article]{achemso}

\usepackage[version=3]{mhchem} 
\usepackage[utf8]{inputenc}
\usepackage[T1]{fontenc}
\usepackage{mathptmx}
\usepackage{etoolbox}
\usepackage{siunitx}
\usepackage{xcolor}
\usepackage[pagewise]{lineno}

\author{A. Malfondet}
 \email{alix.malfondet@u-bourgogne.fr}
\author{M. Deroh}%

\affiliation{ 
Université de Bourgogne, Laboratoire Interdisciplinaire Carnot de Bourgogne (ICB),  CNRS, UMR 6303, F-21078 Dijon, France 
}%

\author{S.-E. Ahmedou}
\affiliation{%
Université de Limoges, CNRS, XLIM, UMR 7252, F-87000 Limoges, France
}%

\author{A. Parriaux}
\affiliation{%
Université de Neuchâtel, Laboratoire Temps-Fréquence, Institut de Physique, Avenue de Bellevaux 51, 2000 Neuchâtel, Switzerland
}%

\author{K. Hammani}
\affiliation{ 
Université de Bourgogne, Laboratoire Interdisciplinaire Carnot de Bourgogne (ICB),  CNRS, UMR 6303, F-21078 Dijon, France 
}%

\author{R. Dauliat}
\affiliation{%
Université de Limoges, CNRS, XLIM, UMR 7252, F-87000 Limoges, France
}%

\author{L. Labonté}
\affiliation{%
Université Côte d’Azur, CNRS, Institut de Physique de Nice, F-06108 Nice Cedex 2, France
}%

\author{S. Tanzilli}
\affiliation{%
Université Côte d’Azur, CNRS, Institut de Physique de Nice, F-06108 Nice Cedex 2, France
}%

\author{J.C. Delagnes}
\affiliation{%
Université de Bordeaux-CNRS-CEA, CELIA, Centre Lasers Intenses et Applications,  UMR 5107, F-
33405 Talence Cedex, France
}%

\author{P. Roy}
\affiliation{%
Université de Limoges, CNRS, XLIM, UMR 7252, F-87000 Limoges, France
}%

\author{R. Jamier}
\affiliation{%
Université de Limoges, CNRS, XLIM, UMR 7252, F-87000 Limoges, France
}%

\author{G. Millot}
\affiliation{ 
Université de Bourgogne, Laboratoire Interdisciplinaire Carnot de Bourgogne (ICB),  CNRS, UMR 6303, F-21078 Dijon, France 
}%
\alsoaffiliation[]
{Institut Universitaire de France (IUF), 1 Rue Descartes, Paris, France}

\title{Near-infrared dual-comb spectroscopy of $\text{C}\text{O}_\text{2}$ and $\text{N}_\text{2}\text{O}$ with a discretized highly nonlinear fiber}

\abbreviations{OFC, DCS, HNLF}

\begin{document}

\begin{tocentry}

Some journals require a graphical entry for the Table of Contents.
This should be laid out ``print ready'' so that the sizing of the
text is correct.

Inside the \texttt{tocentry} environment, the font used is Helvetica
8\,pt, as required by \emph{Journal of the American Chemical
Society}.

The surrounding frame is 9\,cm by 3.5\,cm, which is the maximum
permitted for  \emph{Journal of the American Chemical Society}
graphical table of content entries. The box will not resize if the
content is too big: instead it will overflow the edge of the box.

This box and the associated title will always be printed on a
separate page at the end of the document.

\end{tocentry}

\begin{abstract}
In this paper, we introduce an   all-fiber  dual-comb spectrometer  based on an original design of highly nonlinear fiber to efficiently convert frequency combs in the near infrared from \SI{1.55}{\micro\metre} to around \SI{2}{\micro\metre}. We show that our spectrometer can be used to measure absorption profiles of rovibrational transitions of  $\text{C}\text{O}_2$ and  $\text{N}_2\text{O}$ molecules, and especially their collisional self-broadening coefficients. The results show very good agreement with the HITRAN database and thus further measurements have been performed on a mixture  $\text{C}\text{O}_2$/$\text{N}_2\text{O}$ to measure the broadening of the $\text{C}\text{O}_2$ absorption lines resulting from the presence of $\text{N}_2\text{O}$.
\end{abstract}

\section{Introduction}
Optical frequency combs (OFCs), consisting of evenly spaced and coherent frequency lines within a broad spectrum, are very interesting and powerful tools in a wide range of area including time and frequency metrology, microwaves generation and molecular spectroscopy \cite{hansch_nobel_2006, hall_nobel_2006, jones_carrier-envelope_2000, udem_optical_2002, diddams_direct_2000, cundiff_colloquium_2003, kippenberg_dissipative_2018, picque_frequency_2019}. Over the past decade, numerous technologies and experimental configurations utilizing OFCs for sample interrogation have been extensively demonstrated and reported \cite{keilmann_time-domain_2004, schliesser_frequency-comb_2005, gohle_frequency_2007, coluccelli_optical_2016, thorpe_broadband_2006, coddington_coherent_2008, mandon_fourier_2009, bernhardt_cavity-enhanced_2010}. Among such techniques, dual-comb spectroscopy (DCS), which relies on the interference of two mutually coherent frequency combs that have a slightly different repetition rates, has emerged as a powerful tool enabling real-time detection \cite{schiller_spectrometry_2002, Coddington2016}. DCS finds applications in diverse fields such as breath analysis, agriculture gas flux measurement, and material characterization \cite{Coddington2016, Millot2015, Wang2009, Herman2021, Asahara2016}. It is worthwhile to mention the remarkable technological advancements in optical devices, including electro-optic modulators, fiber couplers, and laser sources operating in the traditional \SI{1.55}{\micro\metre} telecommunication band, which have facilitated the development of dual-comb spectrometers achieving high record performance. However, this spectral range is unsuitable for the detection of several  species  such as carbon dioxide ($\text{C}\text{O}_2$) or nitrious oxide ($\text{N}_2\text{O}$), for which long absorption length are required. To overcome this problem, it is possible to use  a spectrometer in the spectral range around two microns,  where the absorption line strengths are much stronger for these analytes   \cite{gunning_time_2019}.

Nitrous oxide, the third most important greenhouse gas after carbon dioxide  and methane, despite its lower atmospheric presence in terms of mass, possesses  a remarkable global warming potential, being 25 times more potent than $\text{CH}_4$ and 300 times more potent than $\text{C}\text{O}_2$ \cite{kaur_how_2023, song_co-existence_2023}. It also exhibits the longest atmospheric lifetime and contributes significantly to global warming as well as ozone layer depletion. On a global scale, agriculture represents the largest source of $\text{N}_2\text{O}$ emissions, primarily due to the extensive use of nitrogen fertilizers in intensive farming \cite{bell_quantifying_2016}.

In this study, we demonstrate the efficiency of an all-fibered dual-comb spectrometer operating in the  two-micron band. Down-conversion of frequency combs from 1.55 microns to around 2 microns is achieved via degenerate four-wave mixing \cite{Pitois2003, billat_high-power_2014, parriaux_two-micron_2018} in a discretized highly nonlinear fiber (D-HNLF) of original design, which allows precise control of its dispersion parameters \cite{Ahmedou2022}. In fact, the frequency shift induced by four-wave mixing is highly dependent on the values of the second- and fourth-order dispersion coefficients, hence the need for precise control of these two parameters. On the other hand, the D-HNLF effectively reduces fluctuations in the dispersion coefficients along the fiber, thereby enhancing the efficiency of nonlinear conversion. In a previous work, we gave a preliminary result showing the possibility of converting a pulse train from 1562 nm to 1950 nm with such a fiber. In this paper we demonstrate for the first time the possibility of simultaneously converting and spectrally broadening two electro-optic frequency combs. We show that it is then possible to perform dual-comb spectroscopy of different molecular species in the two-micron band. We are particularly interested in the broadening of the absorption spectral lines by collisional processes between molecules. Our experimental results for the collisional self-broadening coefficients of rotationnal-vibrational (rovibrational) transitions of $\text{C}\text{O}_2$ and $\text{N}_2\text{O}$ molecules exhibit excellent agreement with the HITRAN molecular spectroscopic database \cite{gordon_hitran2020_2022}. Furthermore, experimental measurements have been conducted on a $\text{C}\text{O}_2$/$\text{N}_2\text{O}$ mixture to determine the collisional broadening of several lines of $\text{C}\text{O}_2$ due to the presence of $\text{N}_2\text{O}$. Note that the collisional broadening coefficients of the $\text{C}\text{O}_2$/$\text{N}_2\text{O}$ mixture in the spectral band studied are not available in the HITRAN spectroscopic database. Our experimental measurements could therefore contribute to enriching this important database.

\section{Four-wave mixing frequency conversion}

In the literature, several techniques exist to convert OFCs to higher wavelengths, such as difference frequency generation in a periodically poled lithium niobate (PPLN) crystals \cite{Jin2015,Ycas2018,Niu2022} or supercontinuum generation \cite{Ruehl2011,Nader2018,Guo2020}. In this work, we propose to convert frequency combs from the standard  telecommunication band ($\sim$ \SI{1.55}{\micro\metre}) to the near infrared around \SI{2}{\micro\metre} by means of degenerate four-wave mixing (FWM) occurring in a dispersion controlled HNLF. 

\subsection{Principle of degenerate four-wave mixing}

Four-wave mixing is a nonlinear process corresponding to an energy exchange between four waves (two pumps $\omega_{p1}$ and $\omega_{p2}$, a Stokes wave $\omega_s$, and an anti-Stokes wave $\omega_a$). FWM is called degenerate when the two pumps have the same frequency, and non-degenerate otherwise. In this paper we only consider degenerate FWM,
which will be used in our experiments.


The frequency of the Stokes and anti-Stokes waves for which the parametric gain is maximum is given by the following phase-matching condition :

\begin{equation}
\sum_{m=1 }^{\infty} \cfrac{\beta_{2m}\left(\omega_p\right)}{2m !} \Omega_s^{2m}+\gamma P_p=0
\end{equation}
where $\beta_{2m}$ is the dispersion coefficient of order $2m$, $\Omega_s$ the frequency difference between the pump and the generated waves for which the gain is maximum, $\gamma$ the nonlinear coefficient of the fiber, and $P_p$ the pump power. In our case, the zero-dispersion wavelength (ZDW) of the fiber is close to the pump wavelength, so that the dispersion coefficients must be considered up to the fourth order. The phase matching condition can then be written as:

\begin{equation}
\cfrac{\beta_4}{12} \Omega_s^4+\beta_2 \Omega_s^2+2 \gamma P_p=0
\end{equation}

To amplify the signal, the phase mismatch $\Delta\beta$ must thus satisfy:

\begin{equation}
-4 \gamma P_p \leq \Delta \beta \leq 0
\end{equation}

where $\Delta\beta$ is given by:

\begin{equation}
\Delta \beta=2 \beta_2\left(\omega_{\mathrm{p}}-\omega\right)^2+\beta_4\left(\omega_{\mathrm{p}}-\omega\right)^4 / 16
\end{equation}

Finally, the optimal frequency spacing, for which the gain is maximal, and its width $\Delta f$, are then given by:

\begin{equation}
\label{eqn:fopt}
f_{o p t}=\frac{1}{2 \pi} \sqrt{\frac{-2}{\beta_4} \sqrt{9 \beta_2^2-6 \beta_4 \gamma P}-6 \frac{\beta_2}{\beta_4}} \approx \frac{1}{2 \pi} \sqrt{\frac{-12 \beta_2}{\beta_4}}
\end{equation}

\begin{equation}
\label{eqn:deltaf}
\Delta f \approx \frac{\gamma P_p}{2 \pi \beta_2} \sqrt{\frac{-\beta_4}{3 \beta_2}}
\end{equation}

\begin{figure}[htb]
	\centering
	\includegraphics[width=\linewidth]{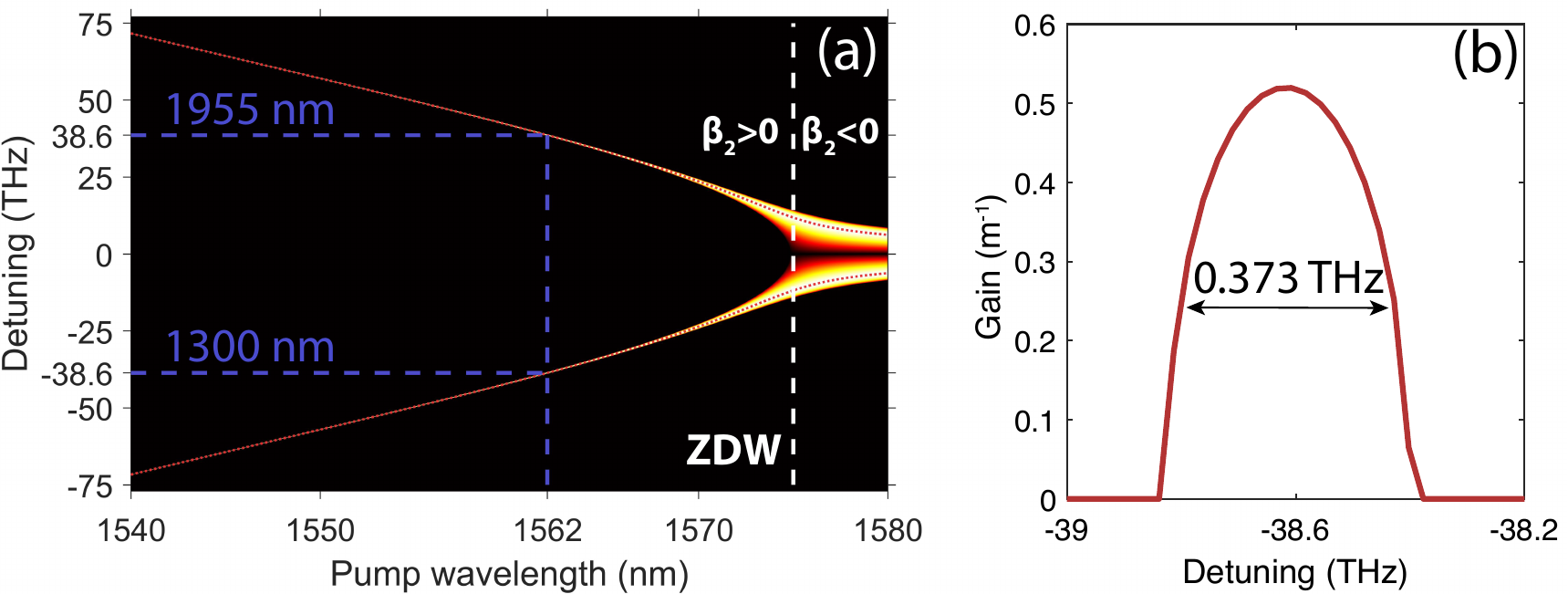}
	\caption{(a) Phase matching diagram for a nonlinear fiber with $\gamma =$ 10 W$^{-1}$.km$^{-1}$ and with $P_p~=~$\SI{26}{\watt}. (b) Gain profile at a pump wavelength of \SI{1962}{\nano \metre}. }
	\label{fig:Phase_matching}
\end{figure}

Equation \ref{eqn:fopt} shows that, to have an efficient conversion at a desired wavelength, the ratio $\beta_2/\beta_4$ has to be controlled precisely. The phase-matching diagram for a discretized nonlinear fiber such as those used below is shown in Fig. \ref{fig:Phase_matching}(a). In the anomalous dispersion regime ($\beta_2 < 0$) we observe a parametric gain characteristic of second-order scalar modulation instability \cite{Agrawal2019}.

On the other hand, in the normal dispersion regime ($\beta_2 > 0$) and with $\beta_4 < 0$, the spectral shift of the parametric gain relative to the pump is much greater, and can reach several tens of THz, leading to idler and signal waves far detuned from the pump. A judicious choice of the dispersion parameters can therefore generate a signal wave in the \SI{2}{\micro \metre} spectral band. If we consider, for example, the case of a pump at \SI{1562}{\nano \metre}, Fig. \ref{fig:Phase_matching}(a) shows that the idler and signal waves are
generated at \SI{1300}{\nano \metre} and \SI{1955}{\nano \metre} respectively. Zooming in at this pump wavelength shows that
the spectral width of the parametric gain is 0.37 THz (Fig. \ref{fig:Phase_matching}(b)), in good agreement with the approximate analytical formula given by Eq. \ref{eqn:deltaf}, which is 0.42 THz.

To this end, we now propose an original fiber design to optimize the conversion in the \SI{2}{\micro\metre} range.

\subsection{Special fiber design for frequency conversion}

\begin{figure}[htb]
	\centering
	\includegraphics[width=\linewidth]{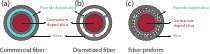}
	\caption{Cross section of a commercial highly nonlinear fiber (a) and of the proposed fiber design (b). Arrangement of the primary D-HNLF fiber preform with three inclusions (c).}
	\label{fig:Principle_3T_V2}
\end{figure}

\begin{figure}[htb]
	\centering
	\includegraphics[width=.95\linewidth]{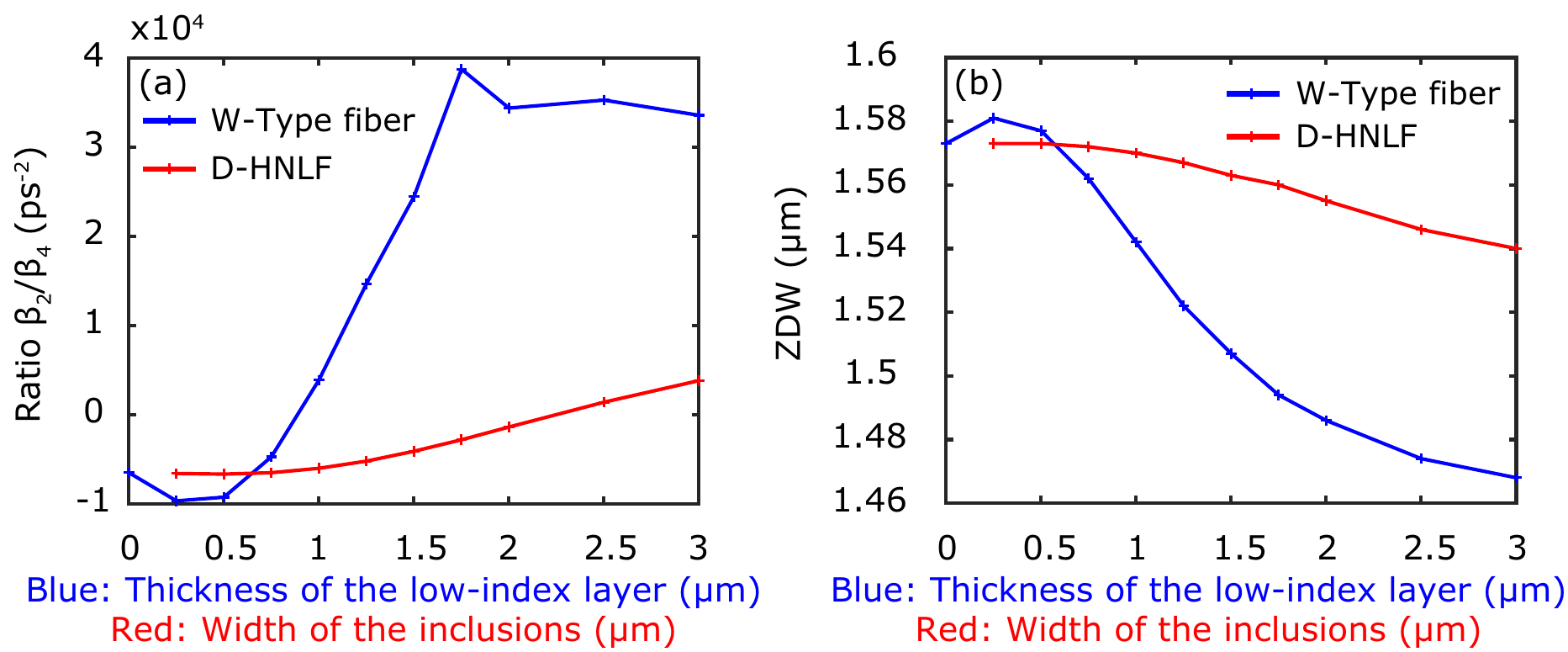}
	\caption{Influence of the thickness of the low-index layer and the width of the inclusions on the ratio $\beta_2/\beta_4$ (a) and on the zero-dispersion wavelength (b).}
	\label{fig:Influence_ZDW_et_dispersion}
\end{figure}

First, let us take the example of a standard commercial W-type HNLF, whose design is depicted in Fig. \ref{fig:Principle_3T_V2} (a). This commercial fiber consists on a germanium doped silica core, surrounded by a low-index layer made of fluorine doped silica. The latter is finally surrounded by a cladding made of pure silica. The drawback of this design is that a small change of its geometry (for instance a change of the thickness of the low-index layer) strongly impacts the dispersion properties of the fiber as shown in Fig. \ref{fig:Influence_ZDW_et_dispersion}. It is thus very difficult to fabricate a HNLF with a precise control of the ZDW and the dispersion coefficients along the fiber, and thus with an efficient conversion at the targeted wavelength \cite{Myslivets2010,Kuo2012}.

To improve the conversion efficiency, we propose to use the design depicted in Fig. \ref{fig:Principle_3T_V2} (b), for which the low-index layer is replaced by low index inclusions of the same material. We show here the example of a discretized optical fiber with three inclusions, as this is the fiber we used in this paper, but we also studied the case of fibers with 2, 4 and 6 inclusions. Our technique consists in discretizing the low-index ring surrounding the core of an optical fiber with a W-index profile (Fig. \ref{fig:Principle_3T_V2} (b)). The low-index inclusions (blue discs in Fig. \ref{fig:Principle_3T_V2}(b)) are exactly juxtaposed to the germanium-doped core zone (red disc in Fig. \ref{fig:Principle_3T_V2}(b)). The discretization of the low-index ring enables finer control of the coupling of the evanescent part of the fundamental mode field with the cladding modes as wavelength increases. So the discretized highly nonlinear fiber (D-HNLF) strongly reduces the constraints imposed on the guided mode of the core, thus allowing a better control of the ZDW and the ratio $\beta_2/\beta_4$ during the fabrication process. The D-HNLF also allows us to reduce longitudinal and transverse fluctuations of the dispersion parameters, which are detrimental for the conversion efficiency\cite{Karlsson1998,Tsuji2002}.

The fabrication  of the D-HNLF is based on the "stack and draw" method (Fig. \ref{fig:Principle_3T_V2} (c)). This consists in drawing, at high temperature, the optical fiber whose outside diameter is on the scale of a hundred \SI{}{\micro\metre}, from an arrangement of capillaries or silica rods forming the fiber structure on a larger scale, typically 10 to 30 mm. The core region was made from a commercial preform with a parabolic index profile. The pure silica and fluorine-doped silica rods are inside a holding tube. The silica ring around the core rod prevents outgassing of the dopants (Ge) during the graded-index preform manufacturing stage. The geometric parameters of the final fiber are obtained by adjusting the drawing conditions: preform descent speed, drawing speed and temperature. A long length of fiber is drawn (500 m) to enable good stabilization of drawing conditions, in particular to obtain thermal stability of the preform in the oven. The chromatic dispersion of the drawn fiber is then measured using a white-light interferometric method (supercontinuum source) based on a conventional Mach-Zehnder interferometer.

To demonstrate the advantages of the D-HNLF, Fig. \ref{fig:Influence_ZDW_et_dispersion} shows the influence of the ZDW (a) and the ratio $\beta_2/\beta_4$ (b) as a function of the thickness of the layer (for the commercial fiber) or the width of the inclusions (for the D-HNLF). The chromatic dispersion is calculated using a commercial software based on the finite element method (COMSOL Multiphysics) giving access to the frequency dependent propagation constant. The simulations depicted in Fig. \ref{fig:Influence_ZDW_et_dispersion} (a) and (b) show that for the commercial fiber (blue line), a small change in the thickness of the low-index layer severely impacts the ZDW and the ratio of the dispersion coefficients. It will therefore be very difficult to optimize such a fiber for a targeted conversion wavelength, because even a small variation of the thickness during fabrication will change the position of the optimal frequency spacing $f_{opt}$. On the other hand, the curves obtained with the D-HNLF (red line) exhibit much smoother slopes. The ZDW and the ratio $\beta_2/\beta_4$ can thus be controlled more easily with this new fiber design, to improve the conversion efficiency at the desired wavelength (i.e. close to \SI{2}{\micro\metre} in our case).

\begin{figure}[htb]
	\centering
	\includegraphics[width=.8\linewidth]{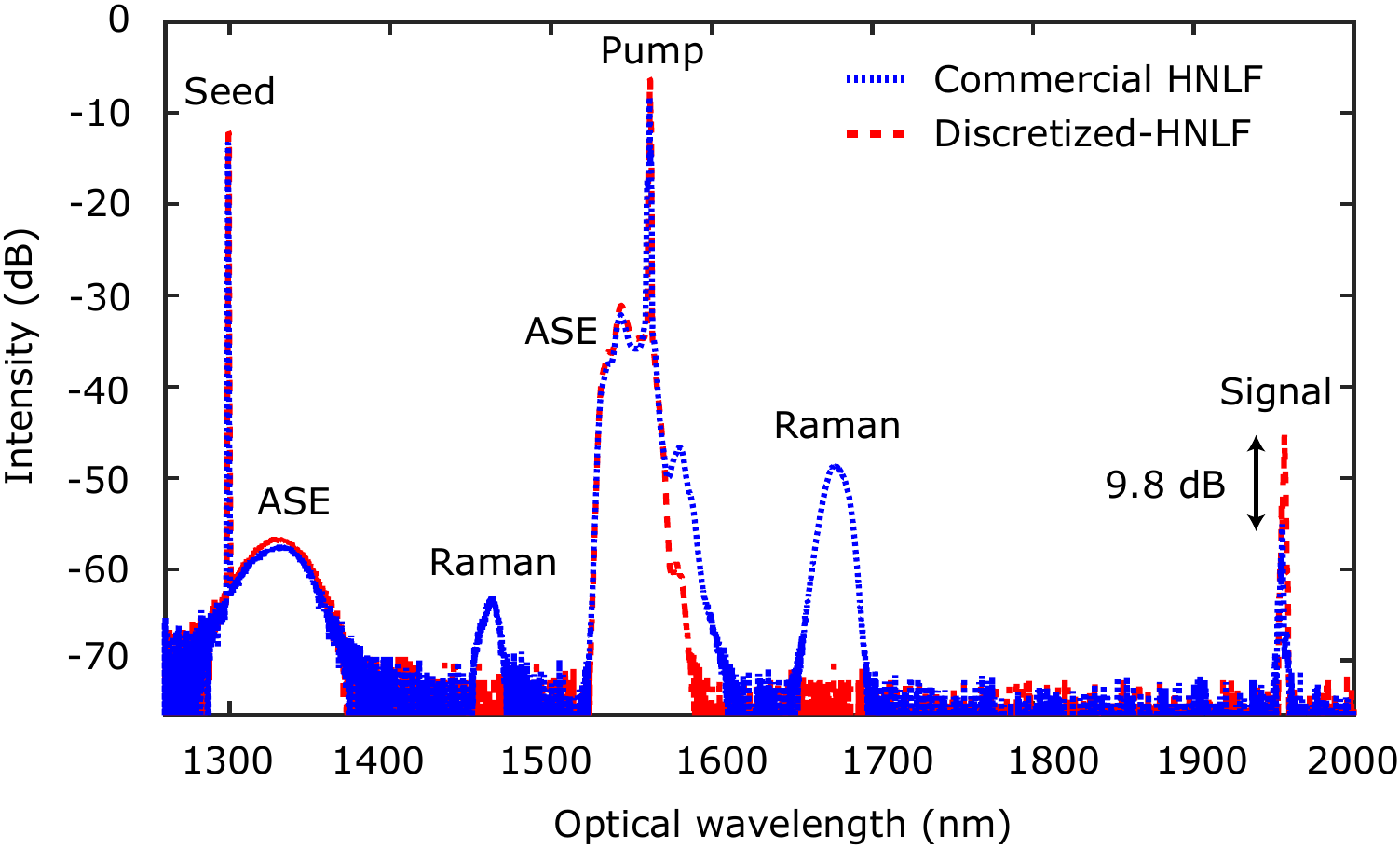}
	\caption{Frequency conversion with a commercial HNLF (blue dots) and with the three-inclusions D-HNLF (red dashed line). The two fibers are 60 m long. $P_{\rm pump}$ = 26 W (peak power) and $P_{\rm seed}$ = 30 mW.}
	\label{fig:Spectre_SUMITOMO_XLIM}
\end{figure}

The improvement due to the D-HNLF is also highlighted in Fig. \ref{fig:Spectre_SUMITOMO_XLIM}, where we compare the spectrum at the output of the D-HNLF (red) and at the output of a commercial HNLF (blue) after injecting a pump and a seed (idler) at their input. To perform these experimental measurements a pump laser source at \SI{1.562}{\micro \metre} is created from a wavelength-tunable continuous source, intensity modulated by an electro-optic modulator, then amplified with an erbium-doped fiber amplifier. Using this technique, we generate a train of 56 ps pulses, measured with an ultra-fast photodiode and a "real-time" oscilloscope, and with a repetition rate of 250 MHz. The pump is then coupled with a continuous wave from a tunable laser diode around \SI{1.3}{\micro \metre} (seed) and amplified via a semiconductor optical amplifier (SOA). The two pump and seed beams are combined via a wavelength multiplexer and then injected into a commercial HNLF fiber (Sumitomo) or one of our D-HNLF fibers. The output of the fiber under test is analyzed using an optical spectrum analyzer to observe the converted signal around \SI{2}{\micro \metre}. For some experiments, a second wavelength multiplexer was inserted just before the spectrum analyzer, to filter out the residual pump and avoid strongly attenuating the signal incident on the optical spectrum analyzer. The various optical fibers tested were all spliced to single-mode optical fiber cords, with measured losses of less than 1 dB per splice. In each experiment, the seed wavelength is finely tuned around 1300 nm to maximize the power of the converted signal near \SI{2}{\micro\metre}.

 Fig. \ref{fig:Spectre_SUMITOMO_XLIM} shows that for  the  seed (idler) at 1300 nm the converted signal measured at 1960 nm is almost \SI{10}{\decibel} higher with the D-HNLF, compared to the one obtained with a commercial HNLF. Moreover, the spectrum at the output of the D-HNLF exhibits no Raman peaks, whereas two Raman peaks are visible when the commercial fiber is used, probably due to a slightly smaller core size and a higher concentration of germanium in the core. However, the pump depletion caused by these peaks is not itself sufficient to explain the improvement of \SI{10}{\decibel}, which is mainly due to the optimization of the dispersion parameters allowed by the D-HNLF. Note that the idler and signal wavelengths for which the parametric conversion is  the  most efficient are in good agreement with the values given by the phase matching condition (see Fig. \ref{fig:Phase_matching}(a)). Similar spectra have been obtained at the output of different D-HNLFs with two, four and six inclusions. As previously explained the pump and idler wavelengths have been optimized so that the signal intensity generated in the 1.9 to \SI{2}{\micro \metre} wavelength range (wavelength range suitable for $\text{C}\text{O}_2$ spectroscopy) is maximal. The spectra recorded with the various D-HNLFs are quite similar, but we observed a slightly higher signal wave for the three-inclusion fiber, which is why we will use this fiber for the spectroscopic applications described in the next section, although the other D-HLNFs could also have been used. Finally, in all experiments, the polarization of the light just before injection into the nonlinear fiber was adjusted using a polarization controller to maximize the power of the converted signal wave at around \SI{2}{\micro\metre}. This polarization optimization is necessary because of the residual birefringence intrinsic to each fiber. None of the nonlinear fibers (HNLF or D-HNLF) are polarization-maintaining.

\begin{figure}[htb]
	\centering
	\includegraphics[width=\linewidth]{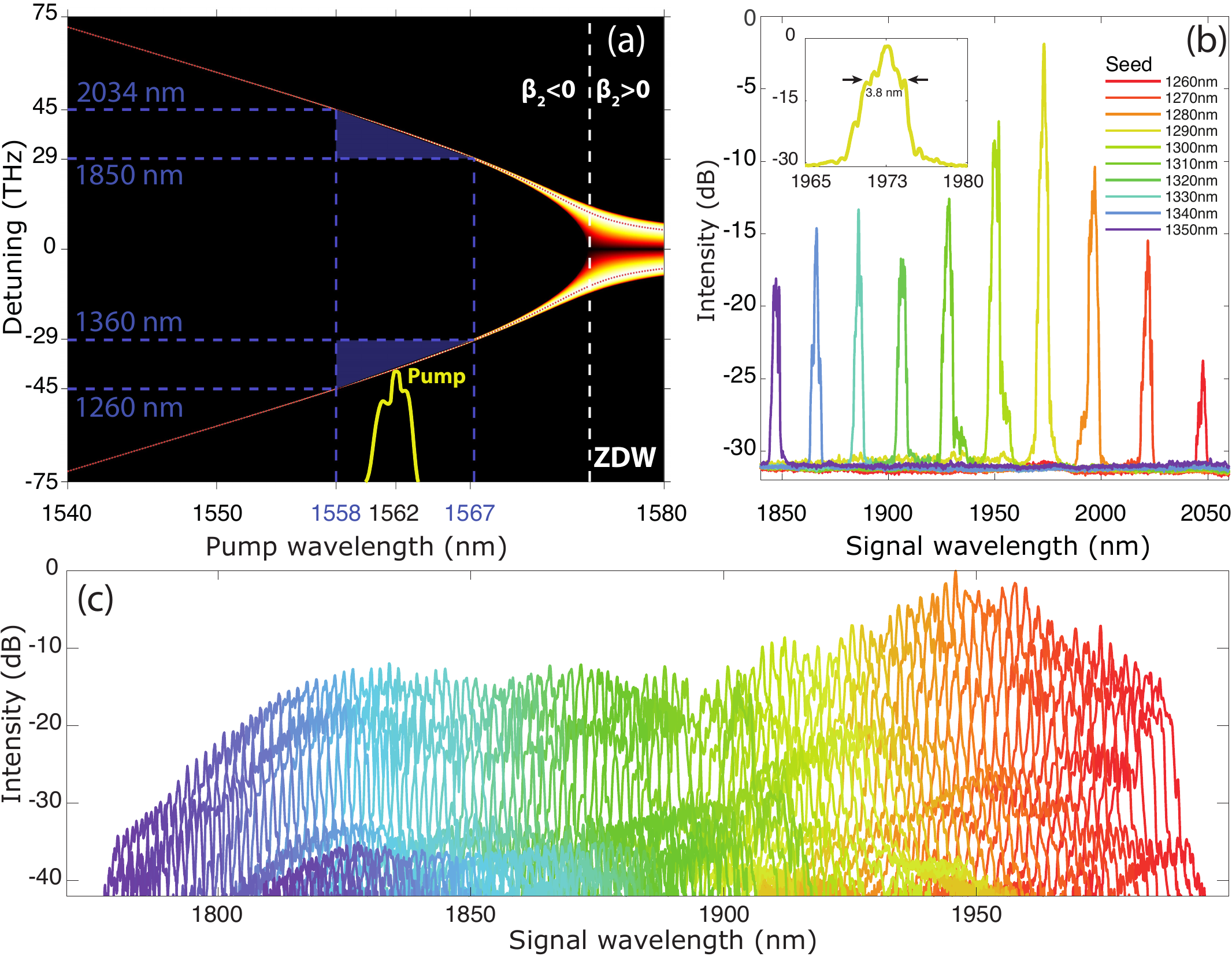}
	\caption{ (a) Phase matching diagram for 3 inclusions fiber with a pumping power  $P_p~=~$\SI{26}{\watt}. (b)~Experimental measurement of the converted signal in a 3 inclusions fiber, obtained by tuning the seed wavelength between \SI{1260}{\nano \metre} and  \SI{1350}{\nano \metre}. The inset correponds to the signal generated using a seed wavelength of  \SI{1290}{\nano \metre}. (c) Experimental measurement of the converted signal in a 2 inclusions fiber, obtained by tuning the seed wavelength between  \SI{1260}{\nano \metre} and \SI{1360}{\nano \metre} using a step of  \SI{1}{\nano \metre}.}
	\label{fig:Largeur_de_conversion}
\end{figure}

Finally, the phase matching diagram of a 3 inclusions fiber, depicted in Fig. \ref{fig:Largeur_de_conversion} shows that it is possible to change the wavelength of the converted signal, by tuning the wavelengths of the pump and the seed signals. Considering our experimental conditions of a tunable seed between \SI{1260}{\nano \metre} and \SI{1360}{\nano \metre}, it is possible to generate a converted signal between \SI{1850}{\nano \metre} and \SI{2034}{\nano \metre}. It is worth noting that the pump wavelength needs to be tuned on a span of only few nanometers, as its -10 dB width is already of the order of  \SI{4}{\nano \metre}, as depicted in the inset of Fig. \ref{fig:Largeur_de_conversion} (b).  The tunability of the frequency conversion is also illustrated in Fig. \ref{fig:Largeur_de_conversion}(b), which shows the experimental converted signal at different seed wavelengths, for a pump wavelength at \SI{1562}{\nano \metre}. It is thus possible to generate a signal between \SI{1850}{\nano \metre} and \SI{2050}{\nano \metre} by tuning the seed wavelength. Another example of a frequency conversion for different seed wavelengths, in a two-inclusions fiber, is also shown in Fig.  \ref{fig:Largeur_de_conversion}(c). The seed is tuned between \SI{1260}{\nano \metre} and \SI{1360}{\nano \metre} with a \SI{1}{\nano \metre} step, allowing to generate a signal between   \SI{1780}{\nano \metre} and \SI{1986}{\nano \metre}.

\color{black}

\section{Demonstration of dual-comb spectroscopy}

\subsection{Experimental setup}

\begin{figure}[htb]
	\centering
	\includegraphics[width=\linewidth]{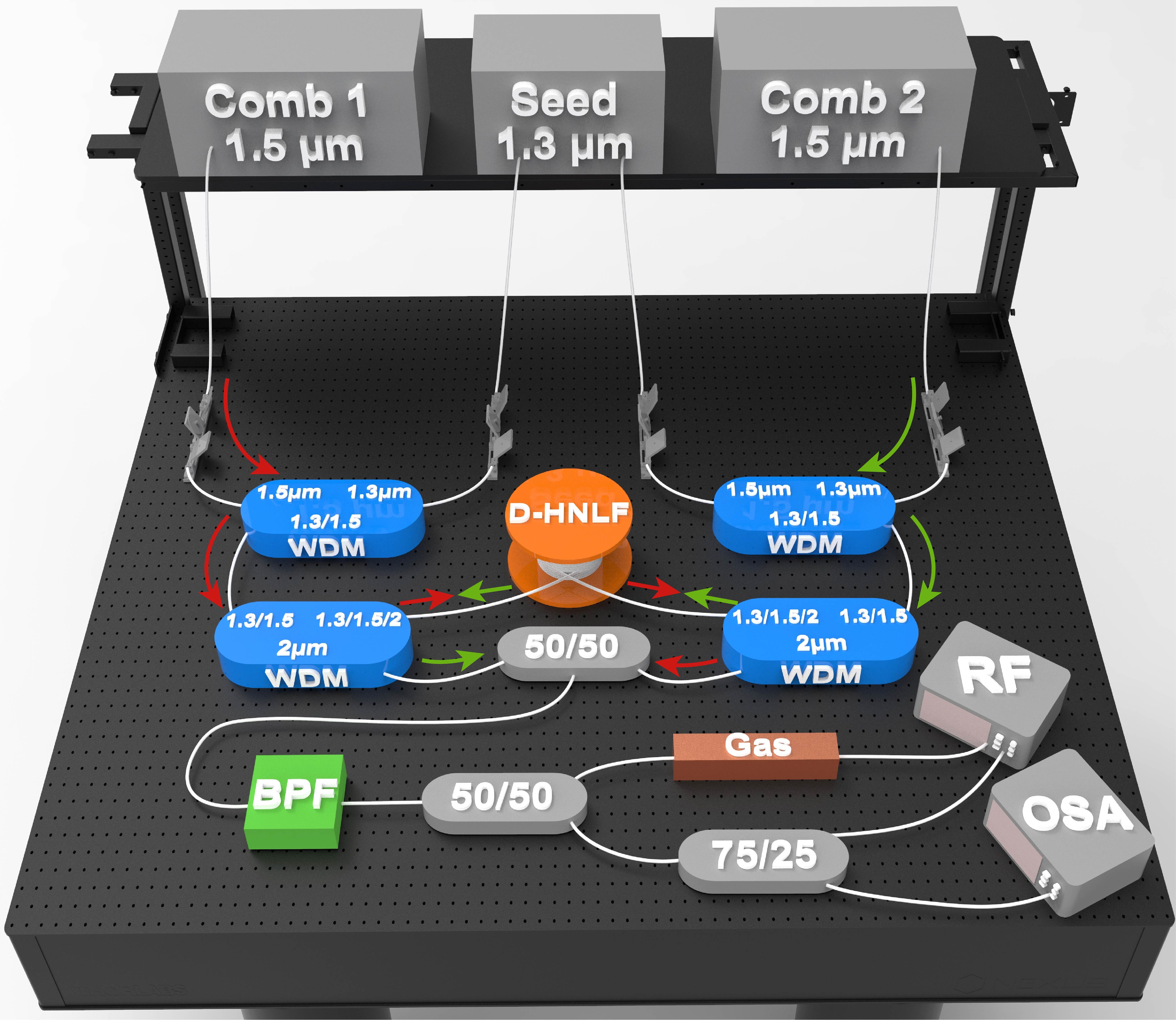}
	\caption{Schematic of the experimental dual-comb spectrometer.}
	\label{fig:Schema_FWM_dual_comb_V1}
\end{figure}

Before presenting the measurements carried out on $\text{CO}_2$ and $\text{N}_2\text{O}$ molecules, we give details of the dual-comb spectrometer  shown in Fig. \ref{fig:Schema_FWM_dual_comb_V1}.

First, two frequency combs (peak power of \SI{43}{\watt}) are generated around \SI{1.55}{\micro\metre} with two electro-optic modulators \cite{parriaux_electro-optic_2020}. The repetition rate of these two combs are respectively $f_{r1}=$\SI{60}{\mega\hertz} and $f_{r2}=f_{r1}+\Delta f_{r}=$\SI{60}{\mega\hertz} + \SI{2}{\kilo\hertz}. The low repetition rate of the combs results in very good spectral resolution, enabling precise analysis of spectral line profiles, particularly at low pressure when the lines are only slightly broadened by the effect of molecular collisions. Each of these two combs are mixed with a \SI{1.3}{\micro\metre} seed (average power of \SI{30}{\milli\watt}) in a 1.3/\SI{1.5}{\micro\metre} wavelength-division multiplexer (WDM), and injected in the 1.3/\SI{1.5}{\micro\metre} port of a second WDM. The two combs and their seed counter-propagate into the 5 meter long D-HNLF, where the \SI{2}{\micro\metre} signals are generated. The D-HNLF is spliced on both sides with standard single mode fiber pigtails, with splicing losses less than \SI{1}{\decibel}. These two signals are then isolated thanks to the \SI{2}{\micro\metre} port of the second WDM, and interfere in a 50/50 coupler. After filtering, the \SI{2}{\micro\metre} interference signal is split into a reference signal (Ref.) that directly goes to the measurement apparatus, and an absorbed signal (Abs.) which goes through a gas cell before being acquired. The Fourier transform of the interference signal gives us the radio frequency (RF) spectrum used to retrieve the absorption profile of the gas under study, whereas the optical spectrum analyzer is used for  wavelength monitoring.

\subsection{Spectroscopy of pure $\text{C}\text{O}_2$ }

\begin{figure}[htb]
	\centering
	\includegraphics[width=\linewidth]{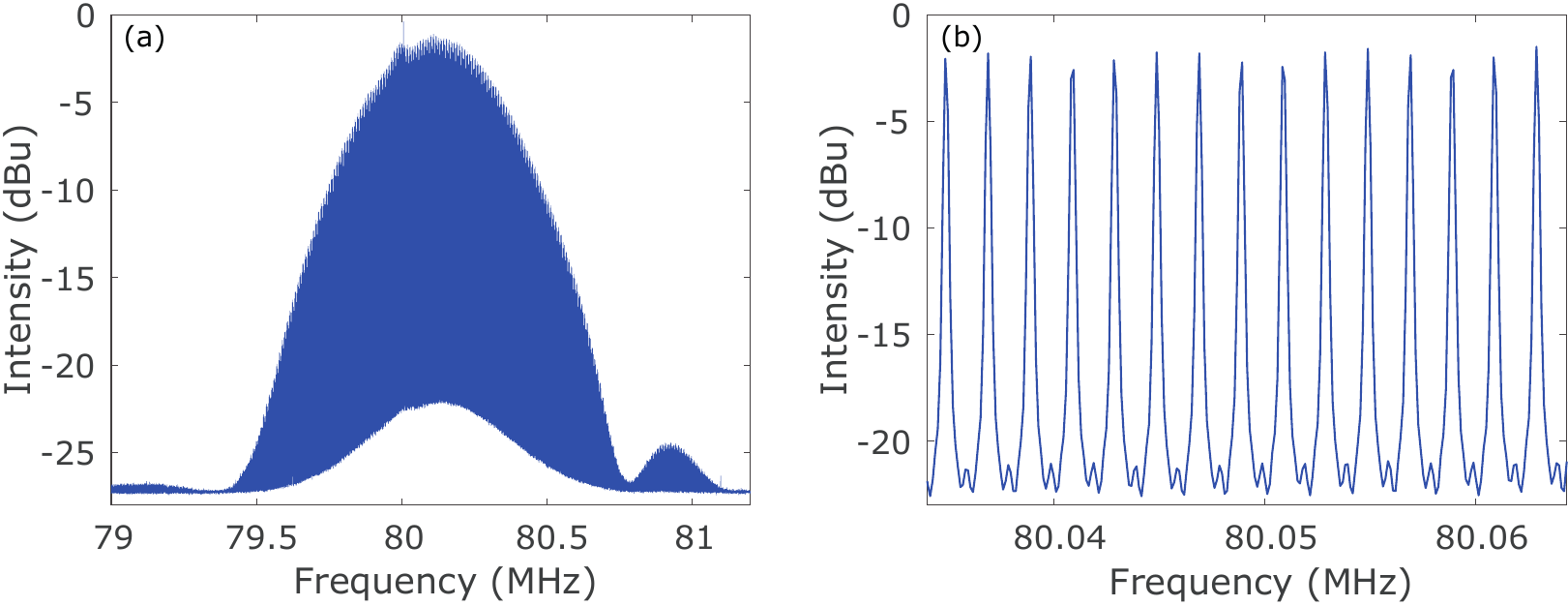}
	\caption{(a): RF spectrum used as a reference. (b): zoom on the RF spectrum.}
	\label{fig:Spectre_RF_1}
\end{figure}

\begin{figure}[htb]
	\centering
	\includegraphics[width=.85\linewidth]{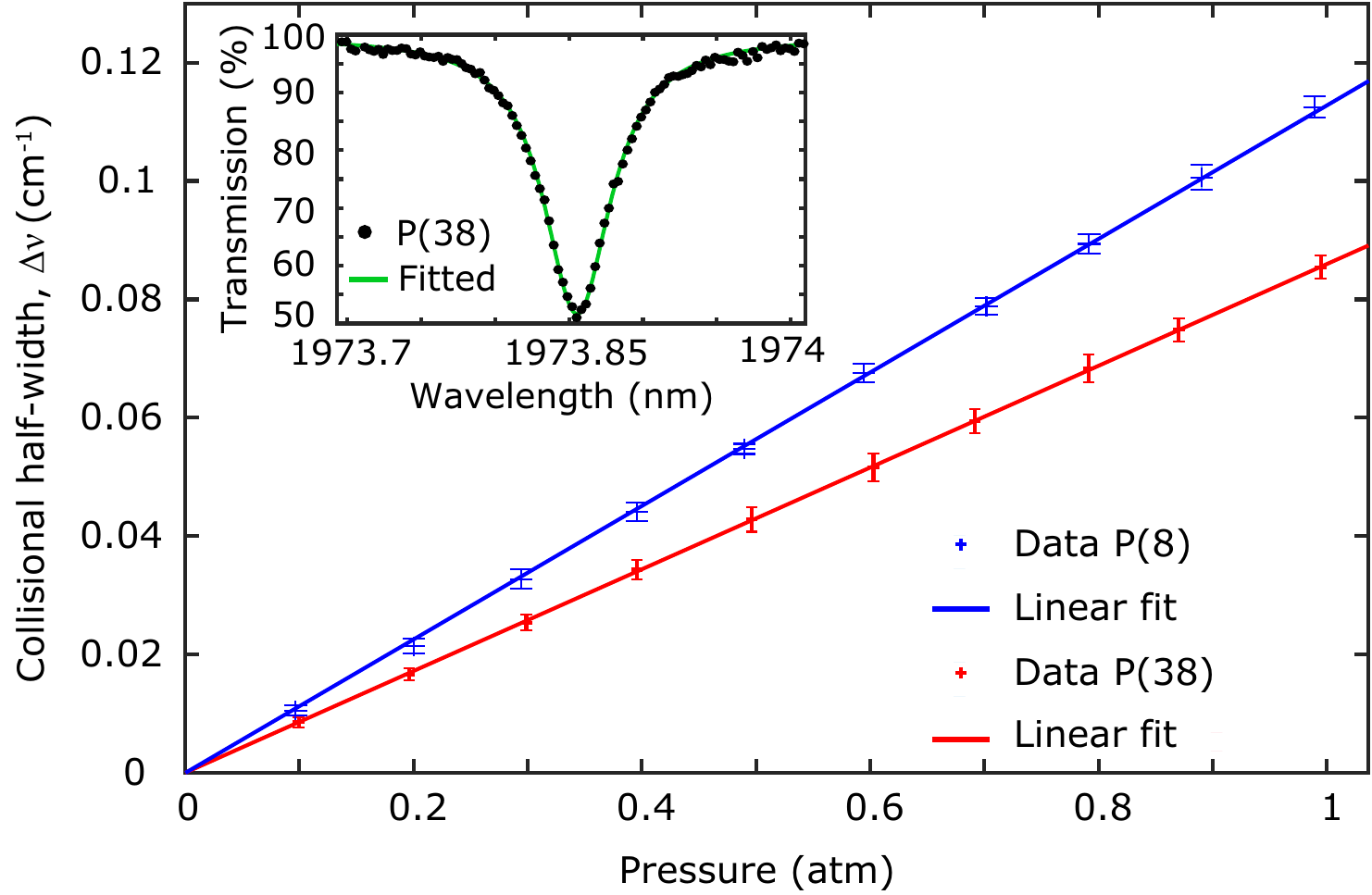}
	\caption{HWHM of the absorption spectral lines of $\text{C}\text{O}_2$, versus pressure, for the P(8) and the P(38) lines. The inset presents an absorption line at a pressure of \SI{700}{\milli\bar} for the P(38) line, measured by the dual-comb spectrometer (black dots) and fitted with a Voigt profile (green line).}
	\label{fig:gamma_CO2_pur}
\end{figure}

We first study the absorption of five spectral lines of pure $\text{C}\text{O}_2$. To this end, the pump and the seed are slightly tuned in order to generate a signal between \SI{1961}{\nano\metre} and \SI{1974}{\nano\metre}. For each spectral line, we measure for different gas pressure, the reference RF spectrum as shown in Fig. \ref{fig:Spectre_RF_1} as well as the absorption spectrum. Finally, the transmission profile (in \%) is obtained by the ratio of the absorption spectrum on the reference spectrum (see the inset of Fig.\ref{fig:gamma_CO2_pur}).

Each transmission profile is compared with a fitted spectrum calculated by the least square regression of a Voigt profile (convolution of a Gaussian profile for Doppler spectral broadening with a Lorentzian profile for collisional spectral broadening) with the parameter lines (frequency, intensity) extracted from the HITRAN database. During the regression, the width of the studied absorption line is adjusted by taking into account the neighboring lines, whose widths are fixed and given by the HITRAN database. From this, we extract the collisional half-width at half-maximum (HWHM) $\Delta\nu$ of the various spectral lines studied as a function of pressure. For clarity, only two of the five rovibrational spectral lines studied with this method are represented in Fig. \ref{fig:gamma_CO2_pur}. The inset in Fig. \ref{fig:gamma_CO2_pur} shows an experimental transmission profile obtained at a pressure of \SI{700}{\milli\bar} (black dots), and compared to the fitted profile (green line). This illustrates the good agreement between our experimental results and the least-squares fitted spectrum. In this way, it is possible to obtain the self-broadening coefficient  $\gamma_{\text{C}\text{O}_2}$ of the rovibrational spectral lines, corresponding to the slope of the curves depicted in Fig. \ref{fig:gamma_CO2_pur}. For the P(8) and the P(38) lines of the $2\nu_1+\nu_3$ band, we obtain collisional self-broadening coefficients respectively equal to ($0.113\pm0.001$) \SI{}{\centi\meter^{-1}}$\text{atm}^{-1}$ and ($0.086\pm0.001$) \SI{}{\centi\meter^{-1}}$\text{atm}^{-1}$, in good agreement with the values obtained with HITRAN, respectively equal to ($0.111\pm0.005$) \SI{}{\centi\meter^{-1}}$\text{atm}^{-1}$ and ($0.082\pm0.004$) \SI{}{\centi\meter^{-1}}$\text{atm}^{-1}$. The experimental uncertainties correspond to twice the standard deviation. Results obtained with the other rovibrational spectral lines are summarized in the table at the last section of this paper.

\subsection{Spectroscopy of pure $\text{N}_2\text{O}$ }
\begin{figure}[htb]
	\centering
	\includegraphics[width=.85\linewidth]{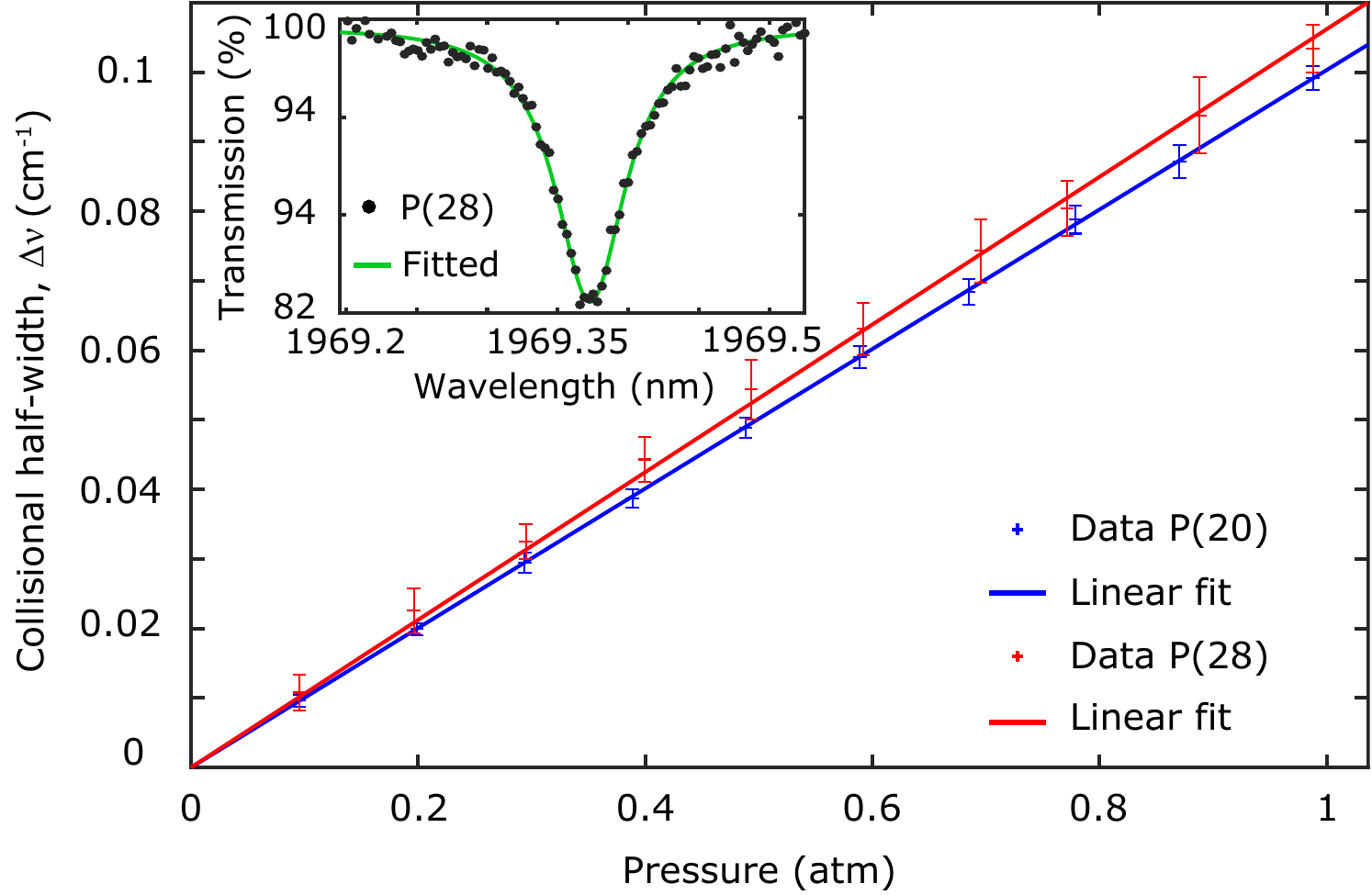}
	\caption{HWHM of the absorption spectral lines of $\text{N}_2\text{O}$, versus pressure, for the P(20) and the P(28) rovibrational spectral lines. The inset presents an absorption line at a pressure of  \SI{700}{\milli\bar} for the P(28) line, measured by the spectrometer (black dots) and fitted with a Voigt profile (green line).}
	\label{fig:gamma_N2O}
\end{figure}

Using the same procedure than for pure $\text{C}\text{O}_2$, we also present results obtained with pure  $\text{N}_2\text{O}$. The seed is now tuned between \SI{1290}{\nano\metre} and \SI{1294}{\nano\metre}, to generate a signal from \SI{1961}{\nano\metre} to \SI{1969}{\nano\metre}. Figure \ref{fig:gamma_N2O} presents the results obtained with the P(20) and the P(28) rovibrational lines. The inset corresponds to an absorption line obtained at \SI{700}{\milli\bar} for the P(28) rovibrational line, where the experimental results (black dots) show a good agreement with the fitted spectrum calculated using the HITRAN database (green line), as explained in the previous subsection.

From the two curves of Fig. \ref{fig:gamma_N2O}, we calculate the self-broadening coefficients $\gamma_{\text{N}_2\text{O}}$ of the P(20) and the P(28) spectral lines of the $4\nu_1$ band, respectively equal to ($0.100\pm0.001$) \SI{}{\centi\meter^{-1}}$\text{atm}^{-1}$ and ($0.106\pm0.002$) \SI{}{\centi\meter^{-1}}$\text{atm}^{-1}$. These values are compatible to those given by HITRAN, respectively equal to ($0.096\pm0.009$) \SI{}{\centi\meter^{-1}}$\text{atm}^{-1}$ and ($0.103\pm0.010$) \SI{}{\centi\meter^{-1}}$\text{atm}^{-1}$.

\subsection{Spectroscopy of a mixture $\text{C}\text{O}_2$/$\text{N}_2\text{O}$ }

\begin{figure}[htb]
	\centering
	\includegraphics[width=.85\linewidth]{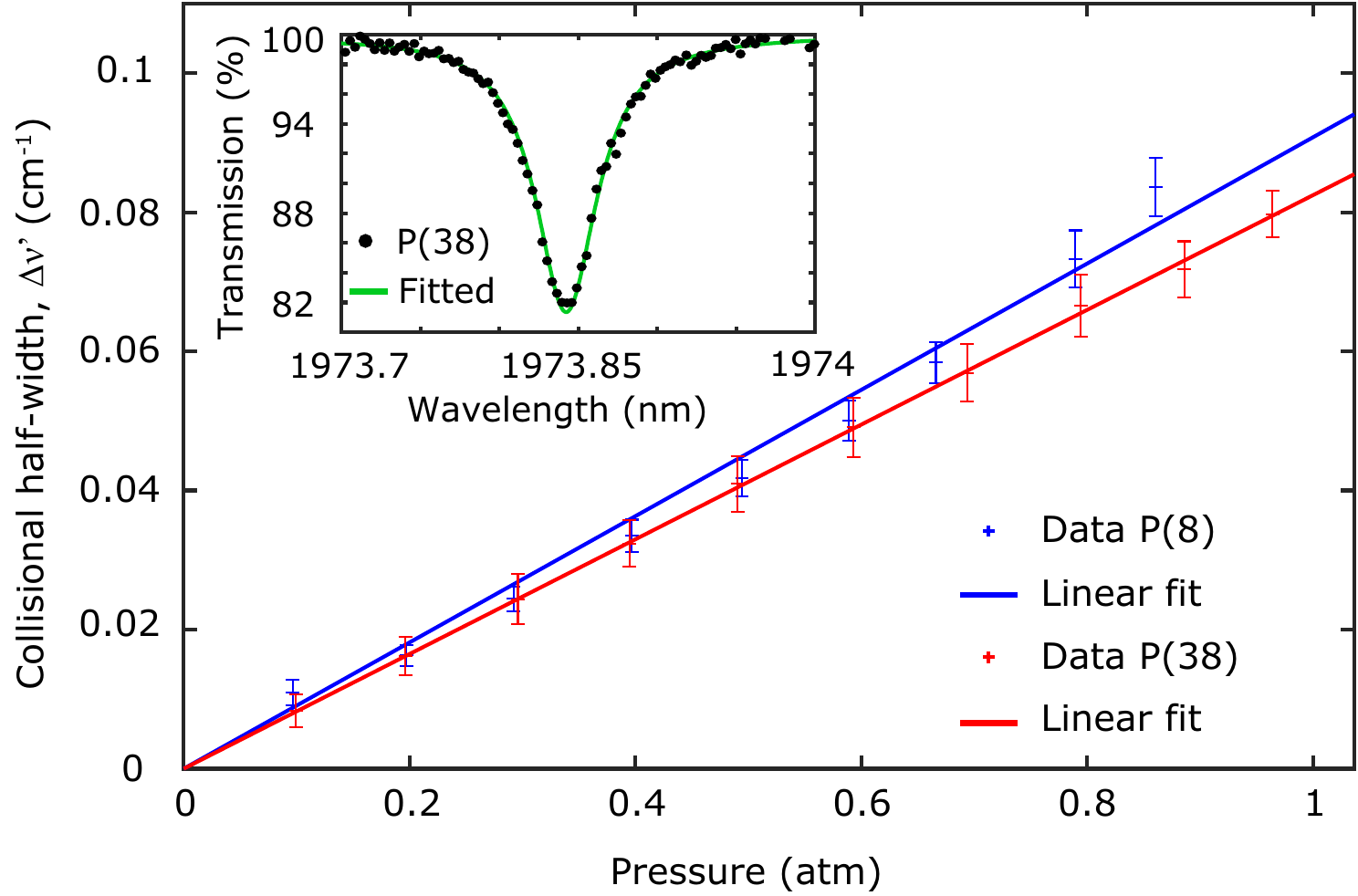}
	\caption{HWHM of the absorption spectral lines of the P(8) and the P(38) rovibrational spectral lines of $\text{C}\text{O}_2$, for a mixture composed of 29.7\% of $\text{C}\text{O}_2$ and 70.3\% of $\text{N}_2\text{O}$. The collisional width is obtained after subtracting the self-broadening contribution. The inset presents an absorption spectral profile at a pressure of \SI{700}{\milli\bar} for the P(38) line, measured by the spectrometer (black dots) and fitted with a Voigt profile (green line).}
	\label{fig:gamma_CO2_N2O}
\end{figure}

We now propose to study a mixture of the two previous gases, composed of $29.7\%$ of $\text{C}\text{O}_2$ and $70.3\%$ of $\text{N}_2\text{O}$. This mixture allows us to measure the broadening coefficient $\gamma_{\text{C}\text{O}_2 / \text{N}_2\text{O}}$, corresponding to the broadening of the rovibrational absorption lines of $\text{C}\text{O}_2$ due to the presence of $\text{N}_2\text{O}$. To this end, we study the same lines that have been studied in the case of pure $\text{C}\text{O}_2$. In case of a mixture $\text{C}\text{O}_2$/$\text{N}_2\text{O}$, the collisional HWHM $\Delta\nu$ of the absorption spectral lines can now be written as :

\begin{equation}
    \Delta\nu = \gamma_{\text{C}\text{O}_2} \text{Pp}_{\text{C}\text{O}_2} + \gamma_{\text{C}\text{O}_2 / \text{N}_2\text{O}} \text{Pp}_{\text{N}_2\text{O}}
\end{equation}

where $\text{Pp}_{\text{C}\text{O}_2}$  and $\text{Pp}_{\text{N}_2\text{O}}$ are respectively equal to  $29.7\%$ and $70.3\%$ of the total pressure. To obtain the broadening coefficient $\gamma_{\text{C}\text{O}_2 / \text{N}_2\text{O}}$, we first subtract the the self-broadening contribution to the width $\Delta\nu$. We obtain a corrected width $\Delta\nu ' = \Delta\nu - \gamma_{\text{C}\text{O}_2} \text{Pp}_{\text{C}\text{O}_2}$, using the value of $\gamma_{\text{C}\text{O}_2}$ that we found earlier. This collisional HWHM is plotted in Fig. \ref{fig:gamma_CO2_N2O} as a function of the partial pressure $\text{Pp}_{\text{N}_2\text{O}}$, to obtain the broadening coefficient $\gamma_{\text{C}\text{O}_2 / \text{N}_2\text{O}}$. The inset in Fig. \ref{fig:gamma_CO2_N2O} shows the good agreement between the absorption lines measured at \SI{700}{\milli\bar} (black dots) and  fitted with a Voigt profile (green line).

The broadening coefficients $\gamma_{\text{C}\text{O}_2 / \text{N}_2\text{O}}$ for the rovibrational lines P(8) and P(38) are respectively found to be equal to ($0.087\pm0.004$) \SI{}{\centi\meter^{-1}}$\text{atm}^{-1}$ and ($0.083\pm0.008$) \SI{}{\centi\meter^{-1}}$\text{atm}^{-1}$.


Finally, we list in the Table  \ref{tab:Summmary_results} the experimental results obtained with the analysis of five absorption rovibrational lines for pure $\text{C}\text{O}_2$, pure $\text{N}_2\text{O}$, and for the mixture $\text{C}\text{O}_2$/$\text{N}_2\text{O}$. For each line, the wavenumber of the seeds and of the generated signals are indicated, as well as the broadening coefficient and its uncertainty for both the experimental measurements and the HITRAN data. These results are compatible with the value given by HITRAN, which confirms the efficiency of the proposed dual-comb spectrometer. It should be noted that the various collisional broadening coefficients are measured with much higher precision than those obtained from the HITRAN database.

\begin{table}[h!]
\centering
\begin{tabular}{l|c|l|l|c|l}
\multicolumn{1}{c|}{\textbf{Parameters}}                                                                                  &  \multicolumn{1}{c|}{\textbf{Pure  $\text{CO}_2$}}                                                      & \multicolumn{1}{c|}{\textbf{Mixture}}                & \multicolumn{1}{c|}{\textbf{Pure $\text{N}_2\text{O}$}}                                                        \\ \hline
\begin{tabular}[c]{@{}l@{}}Line\\ $\nu_s$ ($\SI{}{\centi\meter^{-1}}$)\\ $\gamma_{\text{th.}} \pm \Delta\gamma_{\text{th.}}$ (\SI{}{\centi\meter^{-1}}$\text{atm}^{-1}$)\\ $\gamma_{\text{exp.}}$ $\pm$ $\Delta\gamma_{\text{exp.}}$ (\SI{}{\centi\meter^{-1}}$\text{atm}^{-1}$)\end{tabular} & \begin{tabular}[c]{@{}l@{}} P(2) \\5098.094\\ 0.123$\pm$0.006\\ 0.123$\pm$ 0.001\end{tabular} & \begin{tabular}[c]{@{}l@{}}P(2) \\ 5098.094\\ -\\ 0.095$\pm$0.007\end{tabular} & \begin{tabular}[c]{@{}l@{}}P(8) \\5098.640\\ 0.108$\pm$0.010\\ 0.102$\pm$0.001\end{tabular}   \\ \hline
\begin{tabular}[c]{@{}l@{}}Line \\$\nu_s$ ($\SI{}{\centi\meter^{-1}}$)\\ $\gamma_{\text{th.}} \pm \Delta\gamma_{\text{th.}}$ (\SI{}{\centi\meter^{-1}}$\text{atm}^{-1}$)\\ $\gamma_{\text{exp.}}$ $\pm$ $\Delta\gamma_{\text{exp.}}$ (\SI{}{\centi\meter^{-1}}$\text{atm}^{-1}$)\end{tabular} & \begin{tabular}[c]{@{}l@{}} P(8) \\5093.265\\ 0.111$\pm$0.005\\ 0.113$\pm$0.001\end{tabular}  & \begin{tabular}[c]{@{}l@{}}P(8) \\5093.265\\ -\\ 0.087$\pm$0.004\end{tabular} & \begin{tabular}[c]{@{}l@{}}P(12) \\5094.837\\ 0.103$\pm$0.010\\ 0.102$\pm$0.001\end{tabular}   \\ \hline
\begin{tabular}[c]{@{}l@{}}Line \\$\nu_s$ ($\SI{}{\centi\meter^{-1}}$)\\ $\gamma_{\text{th.}} \pm \Delta\gamma_{\text{th.}}$ (\SI{}{\centi\meter^{-1}}$\text{atm}^{-1}$)\\ $\gamma_{\text{exp.}}$ $\pm$ $\Delta\gamma_{\text{exp.}}$ (\SI{}{\centi\meter^{-1}}$\text{atm}^{-1}$)\end{tabular} & \begin{tabular}[c]{@{}l@{}}P(14) \\5088.242\\ 0.103$\pm$0.005\\ 0.106$\pm$0.003\end{tabular}  & \begin{tabular}[c]{@{}l@{}}P(14) \\5088.242\\ -\\ 0.086$\pm$0.002\end{tabular} & \begin{tabular}[c]{@{}l@{}}P(16) \\5090.846\\ 0.100$\pm$0.010\\ 0.100$\pm$0.001\end{tabular} \\ \hline
\begin{tabular}[c]{@{}l@{}}Line \\$\nu_s$ ($\SI{}{\centi\meter^{-1}}$)\\ $\gamma_{\text{th.}} \pm \Delta\gamma_{\text{th.}}$ (\SI{}{\centi\meter^{-1}}$\text{atm}^{-1}$)\\ $\gamma_{\text{exp.}}$ $\pm$ $\Delta\gamma_{\text{exp.}}$ (\SI{}{\centi\meter^{-1}}$\text{atm}^{-1}$)\end{tabular} & \begin{tabular}[c]{@{}l@{}}P(32) \\5072.043\\ 0.087$\pm$0.004\\ 0.091$\pm$0.001\end{tabular}  & \begin{tabular}[c]{@{}l@{}}P(32)\\ 5072.043\\ -\\ 0.080$\pm$0.002\end{tabular} & \begin{tabular}[c]{@{}l@{}}P(20) \\5086.669\\ 0.096$\pm$0.009\\ 0.100$\pm$0.001\end{tabular}   \\ \hline
\begin{tabular}[c]{@{}l@{}}Line \\$\nu_s$ ($\SI{}{\centi\meter^{-1}}$)\\ $\gamma_{\text{th.}} \pm \Delta\gamma_{\text{th.}}$ (\SI{}{\centi\meter^{-1}}$\text{atm}^{-1}$)\\ $\gamma_{\text{exp.}}$ $\pm$ $\Delta\gamma_{\text{exp.}}$ (\SI{}{\centi\meter^{-1}}$\text{atm}^{-1}$)\end{tabular} & \begin{tabular}[c]{@{}l@{}}P(38) \\5066.283\\ 0.082$\pm$0.004\\ 0.086$\pm$0.001\end{tabular}  & \begin{tabular}[c]{@{}l@{}}P(38) \\5066.283\\ -\\ 0.083$\pm$0.008\end{tabular}   & \begin{tabular}[c]{@{}l@{}}P(28) \\5077.766 \\0.103$\pm$0.010\\ 0.106$\pm$0.002\end{tabular}  
\end{tabular}
\caption{Summary of the experimental results obtained with five absorption spectral lines of pure $\text{CO}_2$, pure  $\text{N}_2\text{O}$, and with the mixture $\text{CO}_2 / \text{N}_2\text{O}$. For each measurement, we indicate the corresponding line, the signal wavenumber $\nu_s$, the broadening coefficient $\gamma_{\text{th.}}$ with its uncertainty $\Delta\gamma_{\text{th.}}$ given by HITRAN, and the experimental broadening coefficient $\gamma_{\text{exp.}}$ with its uncertainty $\Delta\gamma_{\text{exp}}$. The spectral lines of pure $\text{CO}_2$ and of the mixture correspond to the $2\nu_1+\nu_3$ band, whereas the $\text{N}_2\text{O}$ lines correspond to the $4\nu_1$ band.}
\label{tab:Summmary_results}
\end{table}

\clearpage

\section{Conclusion}
To conclude, we have generated frequency combs operating in the near infrared spectral domain, particularly for spectroscopy applications around \SI{2}{\micro\meter}. The method is based on wavelength conversion of frequency combs generated with electro-optic modulators at the telecommunication wavelength near \SI{1.55}{\micro\meter}. The conversion is performed by four-wave mixing process in a special highly nonlinear silica optical fiber characterized by perfectly controlled chromatic dispersion parameters. The original highly nonlinear optical fiber is based on the discretization of a low-index ring surrounding the core with precise and unprecedented management of dispersion coefficients up to fourth order, enabling an order-of-magnitude increase, compared with standard nonlinear commercial fibers, in the conversion efficiency of a signal around \SI{2}{\micro\meter}. This last point is crucial for generating the desired signal, and comes up against problems linked to possible fluctuations in opto-geometric parameters during the fiber manufacturing phase. Remarkably, the original discretized nonlinear fiber simultaneously converts and broadens the frequency combs initially generated at \SI{1.55}{\micro\meter}. After demonstrating the possibility of efficiently converting a comb we show that it is possible to simultaneously convert and broaden two mutually coherent electro-optic combs in a single D-HNLF with a contra-propagating configuration. With the mutual coherence between the two combs preserved, we then proposed an all-fibered dual-comb spectrometer operating in the near-infrared around \SI{2}{\micro\meter} to study the absorption of two important greenhouse gases, namely carbon dioxide $\text{C}\text{O}_2$ and nitrous oxide $\text{N}_2\text{O}$ in both pure and mixed forms. Specifically, we have measured the collisional self-broadening coefficients of a few rovibrational lines, as well as some collisional broadening coefficients due to the presence of $\text{N}_2\text{O}$. The self-broadening coefficients are in excellent agreement with the data provided by the HITRAN molecular spectroscopic database, and we show a much higher accuracy. Collisional broadening coefficients of $\text{C}\text{O}_2$ perturbed by $\text{N}_2\text{O}$ are missing from the large HITRAN database. The precise experimental measurements contribute to the understanding of collisional broadening processes in two gases of environmental interest. Our experimental measurements help to enrich the extensive HITRAN database. Our original nonlinear optical fiber design can be applied to generate frequency combs in other near-infrared spectral regions within the transparency limit of silica glasses. Our concept of discretized nonlinear fibers could also be extended to the mid-infrared with tellurite glasses. The all-fibered dual-comb spectrometer could thus be extended to the mid-infrared, where molecular absorption is very strong.

\section*{Fundings}
The authors would like to thank the Agence Nationale de la Recherche (ANR-17-EURE-0002, ANR-19-CE47-0008, ANR-21-CE42-0026); iXCore Research Foundation; Conseil régional de Bourgogne-Franche-Comté; and the FEDER (Fonds européen de développement régional).

\bibliography{Biblio_article_Dual_Comb_acronyms}

\clearpage

\section{For Table of Contents Use Only}

\begin{figure}[htb]
	\centering
	\includegraphics[width=8.25cm, height=4.45cm]{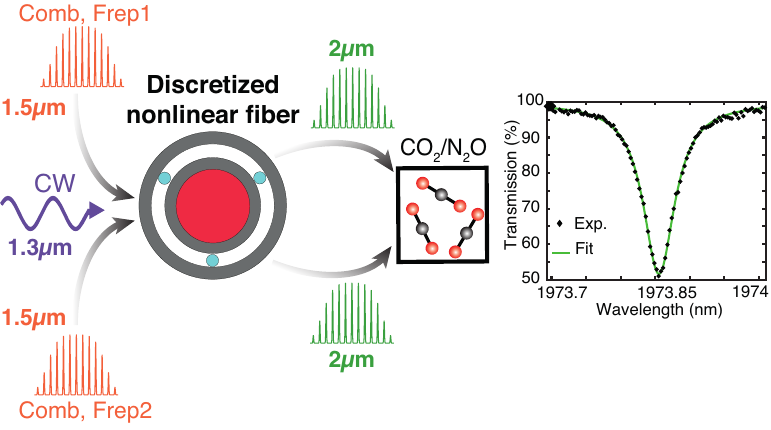}
 \caption{Table of Contents Graphic}
	\label{fig:abstract}
\end{figure}

Near-infrared dual-comb spectroscopy of $\text{C}\text{O}_\text{2}$ and $\text{N}_\text{2}\text{O}$ with a discretized highly nonlinear fiber
\\
\hfill \\
A. Malfondet, M. Deroh, S.-E. Ahmedou, A. Parriaux, K. Hammani, R. Dauliat, L. Labonté, S. Tanzilli, J.C. Delagnes, P. Roy, R. Jamier, and G. Millot\\
\hfill \\
The table of contents graphic depicted in Fig. \ref{fig:abstract} summarizes the experiments presented in this article. Two frequency combs at 1.5 µm wavelength (in orange) and a continuous wave at 1.3 µm (in purple) are injected in a Discretized nonlinear fiber, allowing to generate frequency combs at 2 µm (in green). These frequency combs are then send in a cell filled with  $\text{C}\text{O}_2$ or  $\text{N}_2\text{O}$, in order to analyse these gases. An exprimental measurement of an absorption line is finally shown and compared with a fitted profile.

\end{document}